\documentclass[aps,prb,twocolumn]{revtex4-1}
\usepackage{amssymb}
\usepackage{graphicx}

\begin{document}
\author{Amit and Yogesh Singh }
\affiliation{Department of Physical Sciences, Indian Institute of Science Education and Research (IISER) Mohali, Knowledge City, Sector 81, Mohali 140306, India.}

\date{\today}

\title{Heat capacity evidence for conventional superconductivity in the Type-II Dirac semi-metal PdTe$_2$}

\begin{abstract}
We use electrical transport, magnetoresistance, and heat capacity measurements on high quality single crystals of the recently discovered superconducting Type-II Dirac semi-metal PdTe$_2$, to probe the nature of it's superconducting phase.  The magnitude of the electronic heat capacity anomaly at $T_c$, the low temperature exponential $T$ dependence of the heat capacity, and a conventional $H - T$ phase diagram establish that the superconductivity in PdTe$_2$ is conventional in nature despite the presence of a topologically non-trivial Fermi surface band which contributes to the electrical conduction.   
\end{abstract}

\maketitle  
Topological superconductors have been the focus of intense recent research \cite{Sato}.  This is in part due to the possibility that these materials may host Majorana Fermion excitations \cite{Alicea2012, Elliot2015} which, in addition to being of fundamental interest, can also be used in fault tolerant Quantum computation.  To stabilize topological superconductivity various routes have been pursued.  For example, doping \cite{Hor2010, Kriener2011, Liu2015, Asaba2017, Hor2011, Amit2016, Erickson2009, Balakrishnan2013} or pressurizing \cite{Zhang2011} a parent topological material, studying chiral spin-triplet superconductors \cite{Luke1998}, making heterostructures of a semi-conductor with a conventional superconductor \cite{Sau2010,Alicea2010}, or a topological material with a conventional superconductor \cite{Pribiag2015,Beenakker2016}.  Another exciting new avenue has opened up in which superconductivity has been shown to emerge in nano-scale point contacts between.  Topological materials and normal metals \cite{Aggarwal2016, Wang2016, Das2016}.    

In all these routes, superconductivity is induced by some tuning like doping, pressure, proximity, or confinement.  It would be ideal to look for a system in which Topological band structure and superconductivity emerge naturally and to then demonstrate the Topological character of the superconductivity.  

Recently, a new family of transition metal dichalcogenide materials $AX_2 ~(A =$ Pt, Pd, $X =$ Te, Se) have been shown to be Type-II Dirac materials where the electronic band structure consists of a tilted Dirac cone \cite{H-Huang2016,Noh2017,Fei2017,Yan2017}.  This follows the discovery of Type-II Weyl materials \cite{Soluyanov2015,Weng2015,Xu2015,Deng2016,L-Huang2016,Jiang2017}.  Both the Type-II Weyl and Dirac Fermions observed in the above materials break Lorentz invariance and are therefore fundamentally different quasiparticles compared to the normal Type-I Dirac and Weyl Fermions discovered earlier.  The study of the properties of these Type-II topological materials are therefore of immense fundamental interest and could lead to important technological applications.  How conventional or fairly well understood states of matter like magnetism or superconductivity emerge in materials with Topological band-structures has been an emerging frontier area of research.  In this context, PdTe$_2$ is specially important since it is known to also host a superconducting state below the critical temperature $T_c \approx 1.7$~K~\cite{Jellinek1963}.  Topological superconductivity in PdTe$_2$ is thus an exciting possibility which needs to be carefully examined. 

In this work we report electrical transport, magneto-transport, and heat capacity $C$ measurements on high quality single crystals of the superconducting Type-II Dirac semi-metal PdTe$_2$ to explore the possible unconventional (Topological) nature of the superconducting state.  We confirm superconductivity with a critical temperature $T_c \approx 1.7$~K using electrical transport measurements.  From recent Shubnikov de Haas (SdH) oscillations in magneto-transport measurements we have shown that $4$ bands contribute to the transport, including a band with a non-trivial Berry phase \cite{Das2017}.  This raises the enticing possibility of Topological superconductivity in PdTe$_2$.  Our heat capacity measurements demonstrate bulk superconductivity below $T_c \approx 1.7$~K\@.  The size of the superconducting anomaly in the heat capacity $\Delta C$ at $T_c$ is estimated to be $\Delta C/\gamma Tc \approx 1.52$, which is close to the value $1.43$ expected for a conventional weak-coupling single-gap BCS superconductor.  The $C(T)$ at low temperature shows an exponential $T$ dependence suggesting a fully gapped superconducting state.  Additionally, the $C(T)$ data in various applied magnetic fields $H$ is used to construct an $H$--$T$ phase diagram which also shows a conventional behaviour.   Thus, our measurements strongly indicate that the superconductivity in PdTe$_2$ is conventional in nature despite the presence of Topologically non-trivial electrons contributing to the transport.   

\begin{figure}[t]   
\includegraphics[width= 3 in, angle = 0]{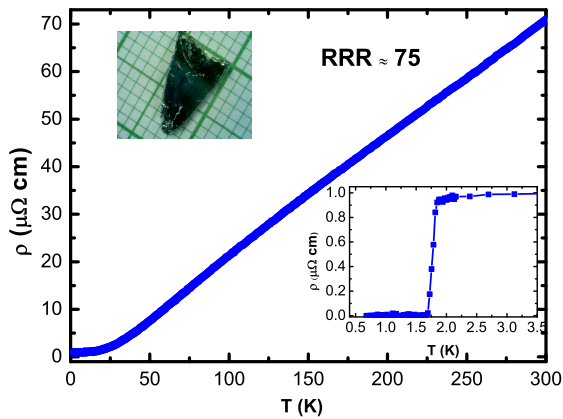}    
\caption{(Color online) Electrical resistivity $\rho$ versus $T$ for PdTe$_2$ measured in zero applied magnetic field with a current $I = 0.5$~mA applied in the crystallographic $ab$-plane.  The top inset shows an optical image of a PdTe$_2$ crystal placed on a milimeter grid.  The bottom inset shows the $\rho(T)$ data below $3$~K to highlight the superconducting transition with $T_c \approx 1.75$~K\@.   
\label{Fig-Res}}
\end{figure}

\noindent
\emph{Experimental:}
Single crystals of PdTe$_2$ were synthesized using a modified Bridgeman method.  The starting elements, Pd powder ($99.99~\%$ purity) and Te shots ($99.9999~\%$), were weighed in the atomic ratio $1 : 2.2$ and sealed in an evacuated quartz tube.  The $10\%$ extra Te was taken to compensate for Te loss due to its high vapor pressure.  For crystal growth, the tube with the starting materials was heated to $790~^o$C in $15$~h, kept there for $48$~h, and then it was slowly cooled to $500~^o$C over $7$ days. They were then annealed at $500~^o$C for $5$~days before cooling naturally. The shiny crystals of millimeter size thus obtained could be cleaved easily from the as grown boule. A typical crystal is shown on a milimeter grid in the inset of Fig.~\ref{Fig-Res}.  The Chemical composition of crystals was verified by energy dispersive spectroscopy (EDS) on a JEOL SEM. The ratio given by EDS between Pd and Te was $1 : 1.99$, showing the stoichiometric ratio of the compound. Few crystals were crushed into powder for X-ray diffraction measurements. The powder X-ray diffraction pattern confirm the phase purity of PdTe$_2$, well crystallized in the CdI$_2$-type structure with the P3m1(164) space group..  

The electrical transport and heat capacity down to $0.4$~K were measured using the He3 option of a quantum design physical property measurement system (QD-PPMS).  

\noindent

\emph{Electrical Transport:}  Figure~\ref{Fig-Res} shows the electrical resistivity $\rho$ versus temperature $T$ measured in zero magnetic field with a current $I = 0.5$~mA applied within the crystallographic $ab$-plane.  The $\rho(T)$ shows metallic behaviour with $\rho(300$~K) $\approx 70~\mu \Omega$~cm and $\rho(2$~K) $\approx 0.94~\mu \Omega$~cm, giving a residual resistivity ratio $RRR \approx 75$.  This $RRR$ is larger than reported earlier indicating that the PdTe$_2$ crystals are of high quality.  The lower inset in Fig.~\ref{Fig-Res} shows the $\rho(T)$ data below $T = 3$~K and the abrupt drop to zero resistance below $T_c = 1.75$~K confirms the superconductivity in PdTe$_2$.

We have recently reported \cite{Das2017} observation of quantum oscillations in the magneto-transport measurements on PdTe$_2$ crystals below $T = 20$~K, again suggesting the high quality of the samples.  For magnetic field applied perpendicular the $c$-axis, we observed a single frequency at $419$~T in the fast Fourier transform of the dHvA data.  The Berry phase for this band was estimated to be non-trivial suggesting its Topological nature.  Additionally, $3$ other frequencies were observed for $H || c$-axis.  Thus there are multiple electronic bands including a Topological band contributing to the transport and it is unclear from just transport measurements whether the observed superconductivity itself has any unconventional Topological character.  


\begin{figure}[t]   
\includegraphics[width= 3 in]{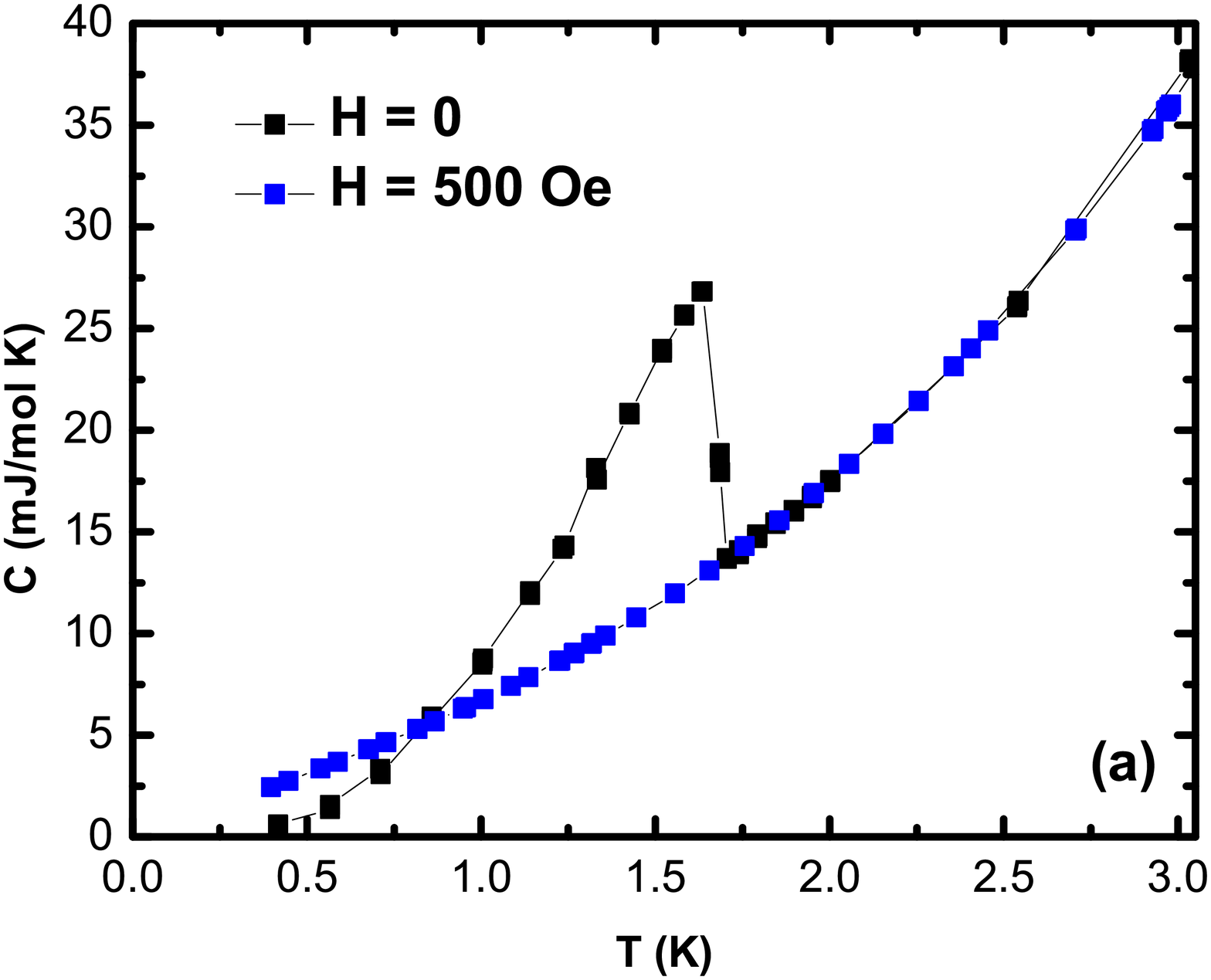}    
\includegraphics[width= 3 in]{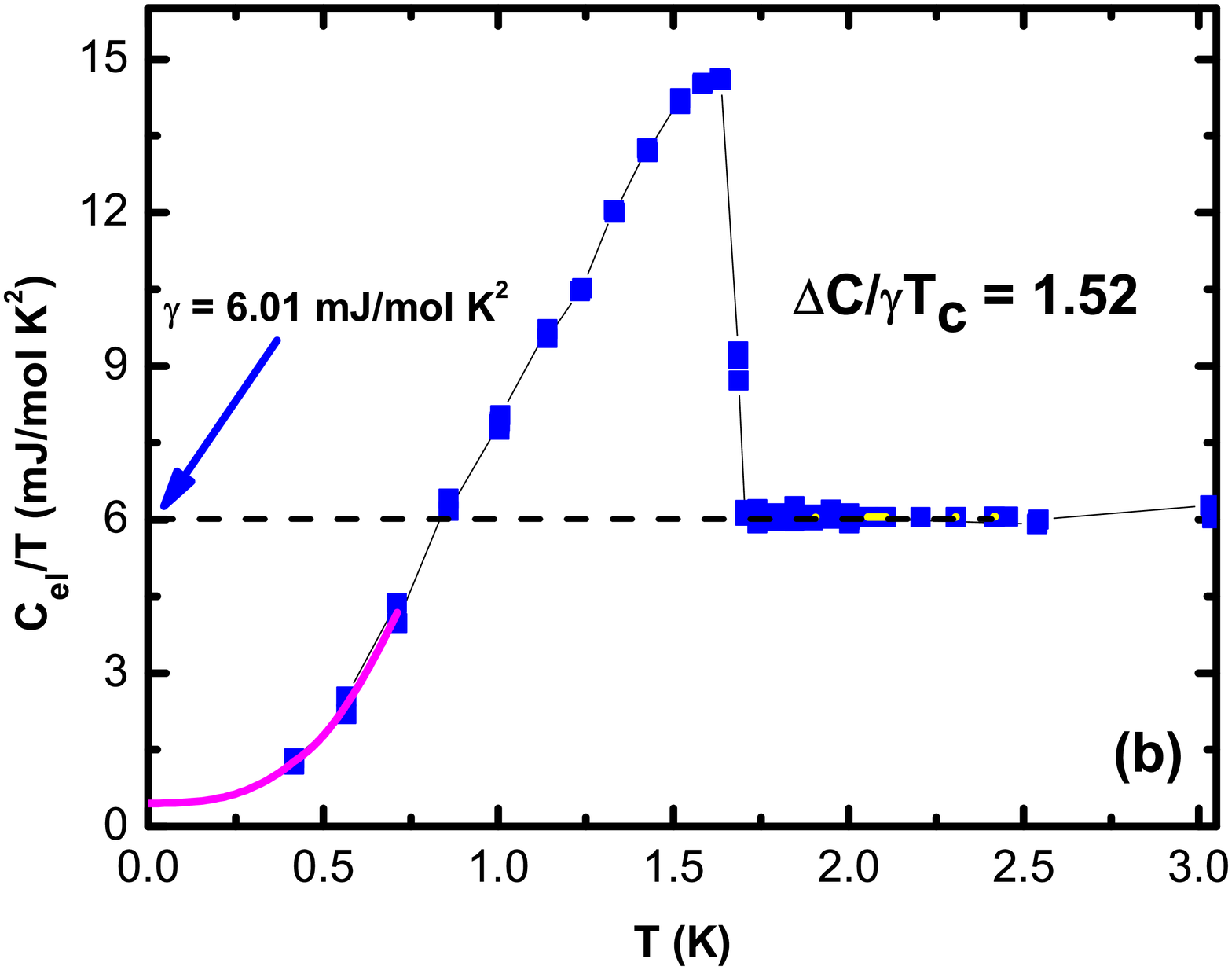}    
\caption{(Color online) (a) Heat capacity $C$ versus $T$ for PdTe$_2$ measured in $H = 0$ and $H = 500$~Oe.  (b) Electronic contribution to the heat capacity divided by temperature $C_{el}/T$. The horizontal dash-dot line is the value $\gamma = 6.01$~mJ/mol~K$^2$ and the solid curve through the lowest $T$ data is a fit by a gapped model (see text for details).    
\label{Fig-C}}
\end{figure}

\begin{figure}[t]   
\includegraphics[width= 3 in]{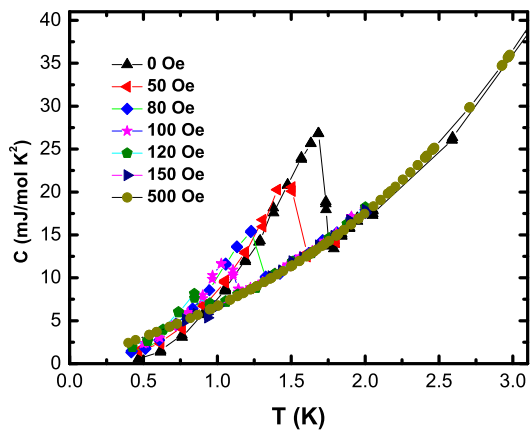}    
\caption{(Color online) Heat capacity $C$ versus $T$ for PdTe$_2$ between $T = 0.4$ and $3$~K, measured in various magnetic fields $H$.
\label{Fig-CH}}
\end{figure}

\noindent
\emph{Heat Capacity:}  We have therefore used heat capacity measurements to address the nature of superconductivity in PdTe$_2$.  Figure~\ref{Fig-C}~(a) show the heat capacity $C$ versus temperature $T$ data for PdTe$_2$ between $T = 0.4$ and $3$~K measured in $H = 0$ and $H = 500$~Oe magnetic field.  A sharp anomaly at $T_c = 1.72$~K in the $H = 0$ data indicates that the superconductivity previously reported using only transport measurements, is bulk in nature.  No anomaly is observed down to the lowest temperatures measured in $H = 500$~Oe, suggesting that the superconductivity has been completely suppressed.  This is confirmed by our heat capacity data in various magnetic field which will be presented later.  The $H = 500$~Oe data was treated as the normal state data and was fit to the expression $C = \gamma T + \beta T^3$.  The fit (not shown) gave the values $\gamma = 6.01(3)$~mJ/mol~K$^2$ and $\beta = 0.66(1)$~mJ/mol~K$^4$.  The lattice part $\beta T^3$ was then subtracted from the $C(T)$ data at $H = 0$ to obtain the electronic part of the heat capacity $C_{el}$.  The electronic heat capacity divided by temperature $C_{el}/T$ versus $T$ is shown in Fig.~\ref{Fig-C}~(b).  An extremely sharp transition at the onset of superconductivity is observed at $T_c = 1.72$~K\@.  The normal state Sommerfeld coefficient $\gamma = 6.01$~mJ/mol~K$^2$ is indicated by an extrapolation (dashed line in Fig.~\ref{Fig-C}~(b)) to $T = 0$ of the normal state data.  An equal entropy construction (not shown) gave almost the same $T_c = 1.69$~K, indicating no broadening or smearing out of the superconducting transition due to sample inhomogeneties or imperfections.  

The data at the lowest temperatures were fit by the expression $C_{el}/T = \gamma_{res} + A exp(-\Delta/T)$, where $\gamma_{res}$ is the residual Sommerfeld coefficient from the non-superconducting fraction of the sample and the second term is a phenomenological exponential decay expected for a gapped ($s$-wave superconductor) system. The fit shown in Fig.~\ref{Fig-C}~(b) as the solid curve through the data below $T = 0.5$~K, gave the value $\gamma_{res} = 0.4$~mJ/mol~K$^2$.  With the total $\gamma = 6.01$~mJ/mol~K$^2$, this suggests that $\approx 7\%$ of the sample volume is non-superconducting.  An excellent fit of the low temperature $C_{el}$ data to an exponential dependence suggests a conventional $s$-wave superconducting order parameter.

The magnitude of the anomaly in heat capacity at the superconducting transition is another measure of the nature (weak or strong coupling, single or multi-gap) of superconductivity.  From the data in Fig.~\ref{Fig-C}~(b) we estimate $\Delta C/\gamma T_c \approx 1.52$, which is close to the value $1.43$ expected for a conventional, weak-coupling, single gap BCS superconductor.  This further supports the conventional nature of superconductivity in PdTe$_2$.

\begin{figure}[t]   
\includegraphics[width= 3 in]{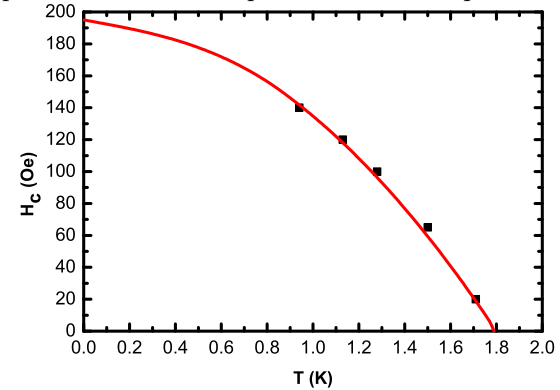}    
\caption{(Color online) The critical magnetic field $H_c$ versus temperature $T$ phase diagram extracted from the $C$ versus $T$ data measured at various $H$ shown in Fig.~\ref{Fig-CH}.  The solid curve through the data is a fit by a phenomenological dependence (se text for details).  
\label{Fig-Hc}}
\end{figure}

Heat capacity measurements at various magnetic fields are shown in Fig.~\ref{Fig-CH}.  As expected, the superconducting transition temperature is monotonically suppressed to lower temperatures and its magnitude becomes smaller at higher fields.  From an equal entropy construction for the $C(T)$ data at each $H$, we extract the $T_c$ at that $H$ and use it to draw a critical-field $H_c$ versus temperature ($H_c-T$) phase diagram.  The $H_c - T$ phase diagram is shown in Fig.~\ref{Fig-Hc} and follows a conventional behaviour expected for a BCS superconductor.  In particular, we were able to fit the data with the phenomenological expression $H_c(T) = H_c(0)[1-(T/T_c)^2]$, with the $T = 0$ critical field $H_c(0)$, and the critical temperature $T_c$ as fit parameters.  The fit shown as the solid curve through the data in Fig.~\ref{Fig-Hc} gave the values $H_c(0) = 195(2)$~Oe and $T_c = 1.78$~K, respectively.  

\noindent
\emph{Summary and Discussion:} PdTe$_2$ is an interesting material where a superconducting state below $T_c \approx 1.7$~K coexists with a Topological band structure. Specifically, PdTe$_2$ has previously been shown to be a Type-II Dirac semi-metal raising the possibility of hosting a Topological superconducting state.  In this study we have used thermodynamic measurements on high quality single crystals of PdTe$_2$ to probe the nature of the superconductivity.  Our heat capacity measurements confirm bulk superconductivity at $T_c = 1.7$~K and show that the anomaly at $T_c$ is characterized by the ratio $\Delta C/\gamma T_c \approx 1.5$ which is close to the value $1.43$ expected for a weak-coupling, single-band BCS superconductor.  The electronic contribution to the heat capacity $C_{el}$ at the lowest temperatures shows an exponential $T$ dependence which points to a gapped $s$-wave superconductivity.  Additionally, the critical field versus temperature phase diagram shows a behaviour expected for a conventional superconductor.  Therefore, all our results strongly indicate that inspite of the presence of a Topological band in the electronic band-structure of PdTe$_2$ which contributes to the transport properties, the superconductivity in PdTe$_2$ is completely conventional and has no Topological character.      

\noindent
\emph{Acknowledgments.--} We thank the X-ray and SEM facilities at IISER Mohali.


\begin{references}

\bibitem{Sato} Masatoshi Sato and Yoichi Ando, Rep. Prog. Phys.,  {\bf 80}, 076501 (2017).

\bibitem{Alicea2012} J. Alicea, Rep. Prog. Phys. {\bf 75}, 076501 (2012).

\bibitem{Elliot2015} S. R. Elliot, M. Franz, Rev. Mod. Phys. {\bf 87}, 137 (2015).

\bibitem{Hor2010} Y. S. Hor, A. J. Williams, J. G. Checkelsky, P. Roushan, J. Seo, Q. Xu, H. W. Zandbergen, A. Yazdani, N. P. Ong, and R. J. Cava, Phys. Rev. Lett. {\bf 104}, 057001 (2010).

\bibitem{Kriener2011} M. Kriener, K. Segawa, Z. Ren, S. Sasaki,. and Y. Ando, Phys. Rev. Lett. {\bf 106}, 127004 (2011).

\bibitem{Liu2015} Z. Liu, J. Am. Chem. Soc. {\bf 137}, 10512 (2015).
 
 \bibitem{Asaba2017} T. Asaba, Phys. Rev. X {\bf 7}, 011009 (2017).
 
\bibitem{Hor2011} Y.S. Hor, J.G. Checkelsky, D. Qu, N. P. Ong, R. J. Cava, Phys. J. Chem. Solids 72, 572 (2011).

\bibitem{Amit2016} Amit and Yogesh Singh, J. Supercond. Novel Mag. {\bf 29}, 1975 (2016).

\bibitem{Erickson2009} A. S. Erickson,  J. -H. Chu, M. F. Toney, T. H. Geballe, and I. R. Fisher, Phys. Rev. B {\bf 79}, 024520 (2009)

\bibitem{Balakrishnan2013} G. Balakrishnan, L. Bawden, S. Cavendish, and M. R. Lees, Phys. Rev. B {\bf 87}, 140507 (2013).

\bibitem{Zhang2011} J. L. Zhang, S. J. Zhang, H. M. Weng, W. Zhang, L. X. Yang, Q. Q. Liu, S. M. Feng, X. C. Wang, R. C. Yu, L. Z. Cao, L. Wang, W. G. Yang, H. Z. Liu, W. Y. Zhao, S. C. Zhang, X. Dai, Z. Fang, and C. Q. Jin, PNAS {\bf 108}, 24 (2011).

\bibitem{Luke1998} G. M. Luke, et al. Nature {\bf 394}, 558 (1998).

\bibitem{Sau2010} Sau et al. Phys. Rev. Lett. {\bf 104}, 040502 (2010).

\bibitem{Alicea2010} J. Alicea Phys. Rev. B {\bf 81}, 125318 (2010).

\bibitem{Pribiag2015} V. S. Pribiag, A.J. A. Beukman, F. Qu, M. C. Cassidy, C. Charpentier, W. Wegscheider, and L. P. Kouwenhoven, Nature Nanotechnology {\bf 10}, 593 (2015).

\bibitem{Beenakker2016} C. Beenakker, and L. Kouwenhoven, Nature Physics {\bf 12}, 618 (2016).

\bibitem{Aggarwal2016} L. Aggarwal, et al. Nature Materials {\bf 15}, 32 (2016).

\bibitem{Wang2016}  H. Wang, H. Wang, H. Liu, H. Lu, W. Yang, S. Jia, X. J. Liu, X. C. Xie, J. Wei and J. Wang, Nature Materials {\bf 15}, 38 (2016).

\bibitem{Das2016} S. Das, L. Aggarwal, S. Roychowdhury, M. Aslam, S. Gayen, K. Biswas, and G. Sheet, Appl. Phys. Lett. {\bf 109}, 132601 (2016).

\bibitem{H-Huang2016} H. Huang, S. Zhou, and W. Duan, Phys. Rev. B {\bf 94}, 121117 (2016).

\bibitem{Noh2017} H. J. Noh, J. Jeong, and E. J. Cho, Phys. Rev. lett. {\bf 119}, 016401 (2017).

\bibitem{Fei2017} F. Fei, et al. Phys. Rev. B. {\bf 96}, 041201 (2017).

\bibitem{Yan2017} M. Yan, et al., Nat. Commun. {\bf 8}, 257 (2017).

\bibitem{Soluyanov2015} A. A. Soluyanov, D. Gresch, Z. Wang, Q. Wu, M. Troyer, X. Dai, and B. A. Bernevig, Nature {\bf 527}, 495 (2015).

\bibitem{Weng2015} H. Weng, C. Fang, Z. Fang, B. A. Bernevig, and X. Dai, Phys. Rev. X {\bf 5}, 011029 (2015).
\bibitem{Xu2015} S.-Y. Xu et al., Science {\bf 349}, 6248 (2015).

\bibitem{Deng2016} K. Deng, et al., Nature Phys. {\bf 12}, 1105 (2016).

\bibitem{L-Huang2016} L. Huang, et al., Nat. Mater. {\bf 15}, 1155 (2016).

\bibitem{Jiang2017} J. Jiang, et al., Nat. Commun. {\bf 8}, 13973 (2017).

\bibitem{Jellinek1963} F. Jellinek, Arkiv. Kemi 20, 447 (1963).

\bibitem{Das2017} S. Das, et al., arXiv:1712.03749v1 (2017).

\bibitem{} T. R. Finlayson, Phys. Rev. B. 33, 2473 (1986).

\end{references}
\end{document}